\documentclass[%
 aip,
 amsmath,amssymb,
 reprint,%
]{revtex4-1}

\usepackage{titlesec}
\titleformat{\section}[block]{\normalfont\bfseries}{\thesection}{1em}{}

\usepackage{graphicx}
\usepackage{dcolumn}
\usepackage{bm}

\usepackage[utf8]{inputenc}
\usepackage[T1]{fontenc}
\usepackage{mathptmx}
\usepackage{etoolbox}

\makeatletter
\def\@email#1#2{%
 \endgroup
 \patchcmd{\titleblock@produce}
  {\frontmatter@RRAPformat}
  {\frontmatter@RRAPformat{\produce@RRAP{*#1\href{mailto:#2}{#2}}}\frontmatter@RRAPformat}
  {}{}
}%
\makeatother
\begin{document}

\preprint{AIP/123-QED}

\title[]{Developing a ChatGPT-Based Tool for Physics Experiment Teaching }
\author{Yifeng Liu}
 \affiliation{Nanchang Institute Of Science and Technology,Nanchang,China}
 
\author{Min Li}%
 \email{cherry198766@126.com}
\affiliation{Nanchang Institute Of Science and Technology,Nanchang,China}

\author{Zhaojun Zhang}
\affiliation{Nanchang Institute Of Science and Technology,Nanchang,China}

\author{Youkang Fang}
\affiliation{Nanchang Institute Of Science and Technology,Nanchang,China}

\author{Meibao Qin}%
 \email{qinmb@ncpu.edu.cn}
\affiliation{Nanchang Institute Of Science and Technology,Nanchang,China}

\date{\today}

\maketitle

\begin{quotation}
In physics experimental teaching, manually writing programs for data analysis and visualization is a common pedagogical method\cite{kozminski2014aapt}. However, it requires specialized skills, is time-consuming, and is prone to human error. As programming skills become increasingly essential in modern physics education, the need for efficient and accurate programming and data processing in physics experiments has emerged as a key challenge. The advent of large language models (LLMs), such as Claude 3.5, has significantly improved language comprehension and code generation capabilities\cite{coello2024effectiveness}. This study explores how tools like Cursor and ChatGPT-4.0-Turbo can assist physics educators without programming experience in developing Python-based tools for synthesizing square wave signals. These tools contribute to enhancing the efficiency and quality of experimental teaching. 
\end{quotation}

\section*{Teaching objectives and content}

The synthesis of square waves provides students with a practical and engaging opportunity to understand the principle of wave superposition, a fundamental concept in physics. Moreover, it serves as an accessible introduction to Fourier analysis, which underpins many critical applications in physics, including acoustics, optics, and circuit analysis. By using the synthesis and analysis of square wave signals as a representative example, the present study aims to bridge theoretical principles with practical teaching applications, making it a valuable resource for physics educators.

The objective of the square wave signal synthesis experiment is to: (1) understand the fundamental characteristics of square wave signals; (2) comprehend the principles of signal synthesis; and (3) master the mathematical principles of Fourier series expansion, and use them to synthesize square wave signals. According to Fourier theory, an ideal symmetric square wave signal can be expressed as an infinite sum of sine waves\cite{James_2011}:

The principle is as follows:

\begin{eqnarray}
I\left( t \right) =\begin{cases}
	anT,&		nT<t<\frac{T}{2}+nT\\
	-\frac{aT}{2},&		\frac{T}{2}+nT<t<\left( n+1 \right) T\quad \left( n\in \mathbb{Z} \right)
\end{cases}
\label{eq:one}.
\end{eqnarray}

The Fourier series decomposition expands 
$I\left( t \right)$ into a series of harmonic signals:

\begin{eqnarray}
I\left( t \right) =\sum_{k=1}^{\infty}{b}_k\sin \left( \frac{2k\pi t}{T} \right) 
\end{eqnarray}

For a square wave signal:

\begin{eqnarray}
b_k=\begin{cases}
	\int_0^{T/2}{a}\sin \left( \frac{2k\pi t}{T} \right) \,dt+\int_{T/2}^T{-}\frac{aT}{2}\sin \left( \frac{2k\pi t}{T} \right) \,dt&		\\
	b_k=\frac{4a}{k\pi}\left( k\text{ for odd harmonics} \right)&		\\
\end{cases}
\end{eqnarray}

That is, the desired square wave can be synthesized using a series of sine waves with amplitudes of $ \frac{4a}{k\pi} $ and frequencies of $ \frac{2k\pi}{T} $:

\begin{eqnarray}
f_{sq}(t) =\sum_{k=1}^{\infty}{\left( \frac{4a}{k\pi}\sin \left( \frac{2k\pi t}{T} \right) \right)},\quad k\text{ for odd harmonics}
\end{eqnarray}

\begin{eqnarray}
f_{sq}(t) =\frac{4a}{\pi}\left( \sin \omega t+\frac{1}{3}\sin 3\omega t+\frac{1}{5}\sin 5\omega t+\cdots \right) 
\end{eqnarray}

where $f_{sq}(t)$ represents the instantaneous value of the ideal square wave at time $t$ (assumed to switch between +1 and –1); $\omega_0$ denotes the angular frequency of the fundamental component of the square wave; $n$ refers to the odd harmonics (i.e., $n=1,3,5,...$). This equation forms the theoretical basis of the square wave synthesis experiment. By successively superimposing sine waves of different frequencies and amplitudes, one can progressively construct a square wave signal.

In this paper, the synthesis of square waves involves combining a series of sine waves generated by the Fourier decomposition synthesizer in an adder.

\section*{Experimental setup and operational challenges}
The principle of square wave signal synthesis in the experiment is shown in Figure 1(a), where a series of sine waves generated by the waveform generator are input into an adder to synthesize the signal, and the waveform is ultimately displayed by the oscilloscope. The experimental wiring diagram is shown in Figure 1(b).

\begin{figure}[htbp]
\includegraphics[scale=0.35]{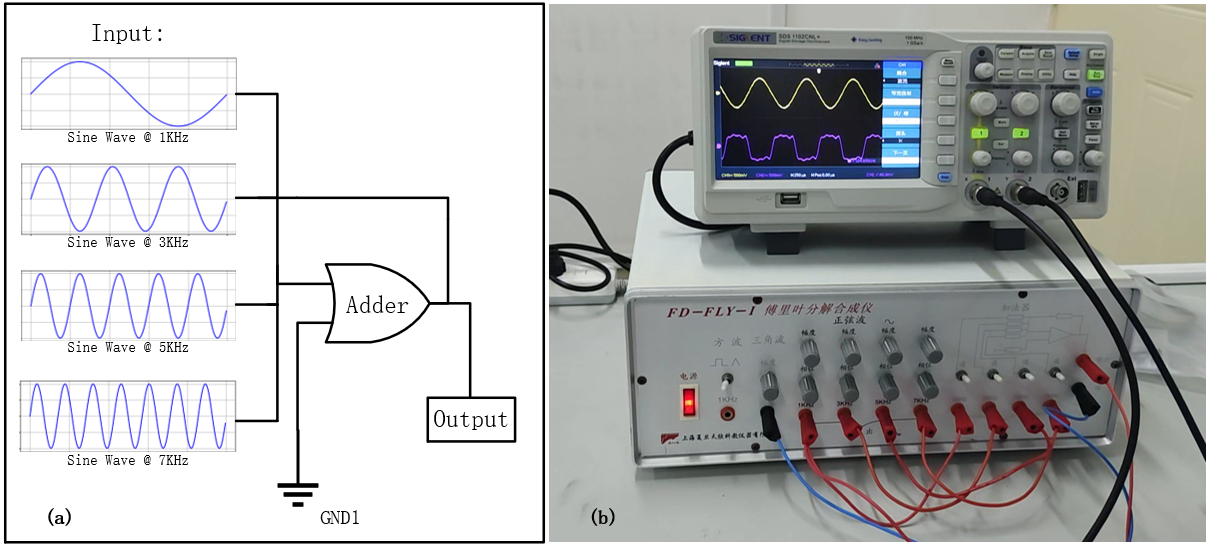}
\caption{\label{fig:epsart} (a) Schematic diagram of the experimental principle.
(b) Completed experimental setup and the synthesized square wave pattern. During the experiment, students are required to independently adjust the amplitude and phase of the sine waves.}
\end{figure}
In the actual experimental operations, the participating students reported the following issues:\\
(1) The experimental procedures, including wiring and instrument calibration, are relatively complex, making the operation quite challenging.\\
(2) The real-world operational environment introduces various influencing factors, leading to unstable waveforms, which makes it difficult to synthesize an ideal square wave pattern. The overall difficulty of the experiment is relatively high, and the results are usually not satisfactory.\\
When problems occur during the experiment or environmental interference affects the results, it is often difficult to identify the causes. If interactive demonstrations are conducted using teaching tools before the experiment, the teaching efficiency can be significantly increased. This tool enables physics educators to offer a visualized and variable-controllable theoretical simulation environment, allowing students to focus on investigating physical processes and analyzing results without being hindered by experimental artifacts or complex procedures.
\section*{Automatically generate experimental programs using ChatGPT}

With the help of ChatGPT and carefully designed prompts, teachers can quickly obtain executable source code. Below are some typical GPT prompts and the corresponding code generated (ChatGPT's replies are not fully presented here, with [...] used to indicate omissions)\\

\textbf{\textit{Input:}} You are a university physics teacher, currently conducting a "Square Wave Synthesis" physics experiment. I need you to write a Python program to demonstrate the synthesis of square waves for the students in class. Additionally, I will send you the course materials. (The course materials are uploaded in the attachment)

\textbf{\textit{ChatGPT:}} Based on the experimental content in the course materials, I will[…]\\

The code was run using PyCharm 2024.3.1.1, and the result is shown in Figure 2. Of course, we can also use some online IDEs (such as \href{URL}{codesandbox.io}
/).\\

\begin{figure}[htbp]
\includegraphics[scale=0.2]{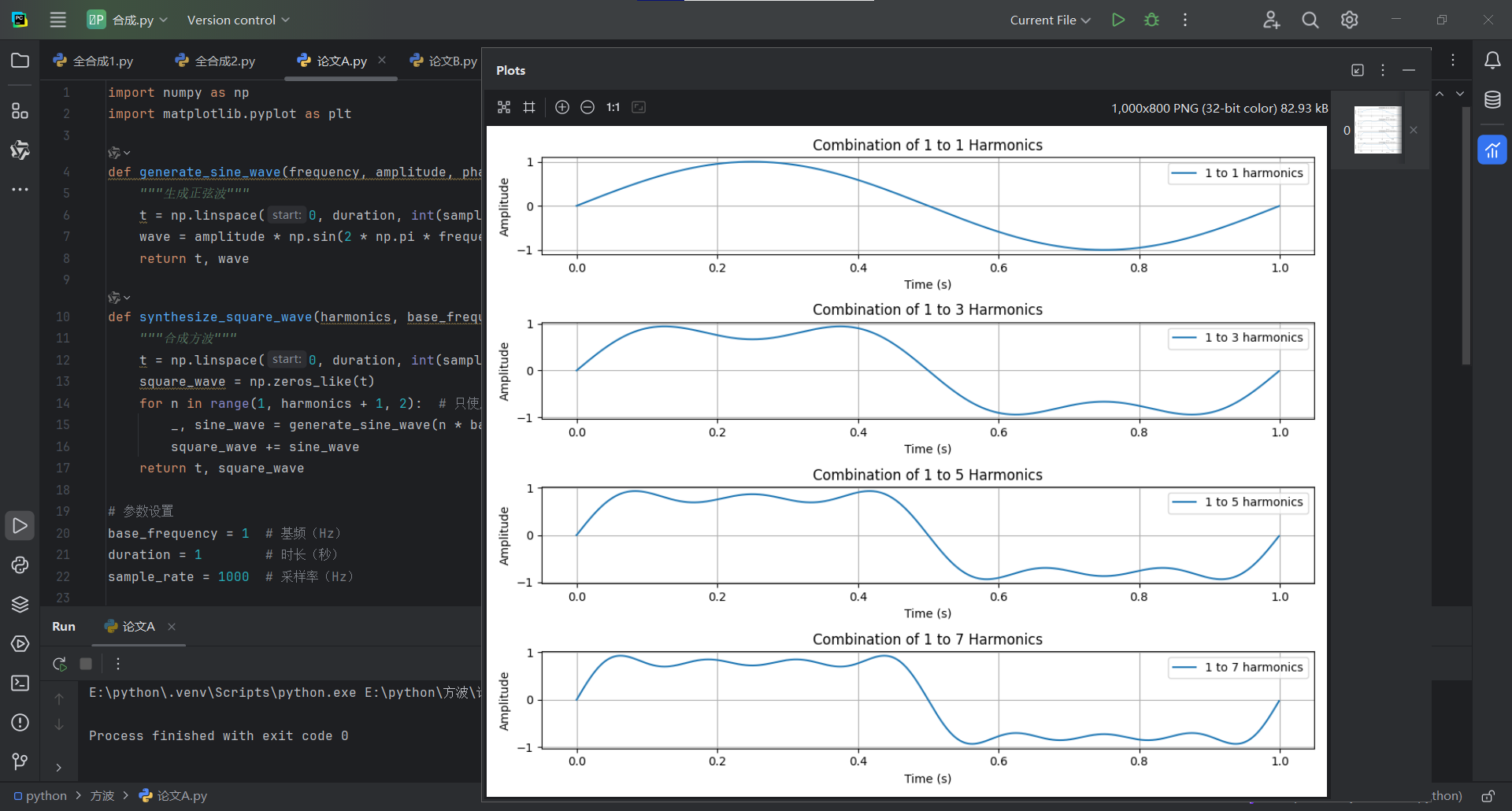}
\caption{\label{fig:epsart} Copy the code and run the result in PyCharm 2024.3.1.1.}
\end{figure}

With the appropriate prompts, we can ask ChatGPT to further optimize the code:\\

\textbf{\textit{Input:}} I don't think this is a perfect program; it only generates an image. Could you add some buttons to display each step of the waveform synthesis process? This would help students better understand the entire signal synthesis process.

\textbf{\textit{ChatGPT:}} Based on your request[…]\\

The result is shown in Fig.3.

\begin{figure}[htbp]
\includegraphics[scale=0.4]{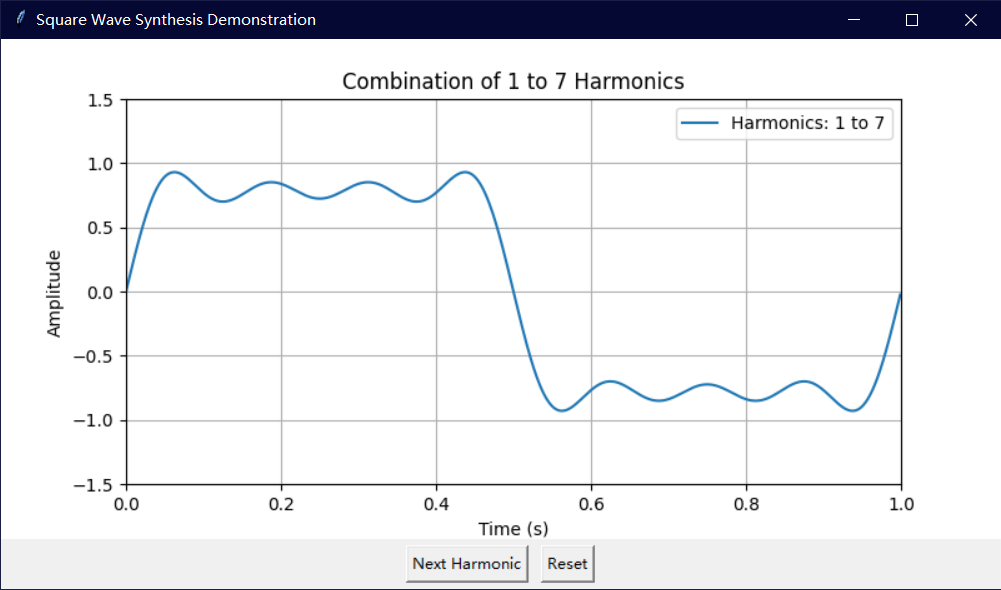}
\caption{\label{fig:epsart} GPT has generated an interactive program for us, with the following functions: \textbf{Next Harmonic:} incrementally adds one harmonic, \textbf{Reset:} resets the number of harmonics to the fundamental (1st order) and redisplays the waveform.}
\end{figure}

After obtaining a satisfactory initial version of the code, it can be imported into the Cursor platform for further debugging and optimization. Finally, the finalized version can be packaged into an EXE program (Fig. 4) using the PyInstaller library\cite{cortesi2022pyinstaller}. This way, even on any Windows computer without a Python environment installed, both teachers and students can run the program and conduct the simulation experiment smoothly.
\begin{figure}[htbp]
\includegraphics[scale=0.3]{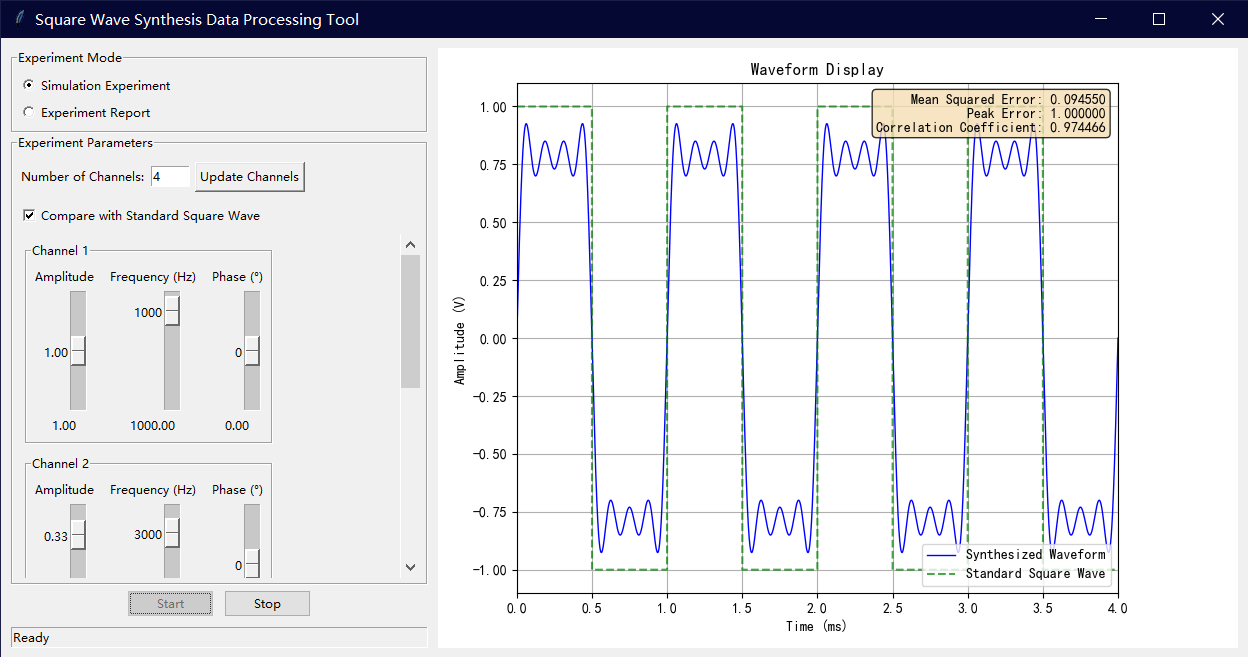}
\caption{\label{fig:epsart} The final square wave synthesis experiment tool EXE program allows real-time adjustment of the number of sine wave channels, the amplitude, frequency, and phase of each channel. Additionally, it can display various error metrics compared to the standard square wave graph in the top-right corner.}
\end{figure}

\section*{Summary and future outlook}

Artificial intelligence (AI) is increasingly being applied in the fields of education and research, particularly in physics experimental teaching, where it demonstrates tremendous potential\cite{hwang2020vision}. This paper presents an example of square wave signal synthesis experiments to showcase the efficiency and practicality of LLMs like ChatGPT in creating experimental tools. With these tools, even physics teachers with no programming experience can easily generate fully functional physics teaching tools, thereby improving teaching efficiency and optimizing students' experimental experience. Therefore, this study has high generalizability. For example, in mechanics, teachers could develop a simple harmonic motion simulator that allows adjustment of physical parameters such as mass, spring constant, or pendulum length to explore phenomena like damping and resonance. In optics, interactive simulations of Young’s double-slit interference or single-slit diffraction could be created, where variables such as wavelength, slit spacing, or slit width can be modified to observe changes in the interference or diffraction patterns in real time, thereby deepening students’ understanding of wave optics. The LLM-assisted development of personalized and interactive teaching tools holds significant promise as an innovative direction in physics education.

\begin{acknowledgments}
This work is supported by the National Natural Science Foundation of China (Grants No. 12364049),  the Natural Science Foundation of Jiangxi Province (Grants No. 20242BAB25041), the Start-up Funding of Nanchang Institute of Science and Technology (Grant No. NGRCZX-23–01), and the Scientific research project of  Nanchang  Institute of Science and Technology (Grant No. NGKJ-23-02).
\end{acknowledgments}

\section*{References}

\nocite{*}
\bibliographystyle{unsrt}
\bibliography{aipsamp}

\end{document}